\documentstyle{article}
\begin{document}
\begin{center}
{\Large\bf The numerical accuracy of truncated Ewald sums for
periodic systems with long-range Coulomb interactions}
\\[0.5\baselineskip]
{\large Gerhard Hummer}\\[0.5\baselineskip]
{Theoretical Biology and Biophysics Group T-10, MS K710,\\
Los Alamos National Laboratory,
Los Alamos, NM 87545, U.S.A.}\\[0.5\baselineskip]
(Chemical Physics Letters: in press, 1995)
\end{center}
\begin{abstract}
Ewald summation is widely used to calculate electrostatic interactions
in computer simulations of condensed-matter systems. We present an
analysis of the errors arising from truncating the infinite real- and
Fourier-space lattice sums in the Ewald formulation. We derive an
optimal choice for the Fourier-space cutoff given a screening parameter
$\eta$. We find that the number of vectors in Fourier space required
to achieve a given accuracy scales with $\eta^3$. The proposed method
can be used to determine computationally
efficient parameters for Ewald sums, to assess
the quality of Ewald-sum implementations, and to compare different
implementations.
\end{abstract}
\section{Introduction}
The calculation of electrostatic interactions in computer-simulation
studies of condensed-matter systems poses serious problems regarding
accuracy and efficiency \cite{Allen:87}. These are mainly caused by the
infinite range of Coulomb interactions in conjunction with the finite
size of the samples studied. To avoid system-size effects,
periodic boundary conditions are employed. The natural description of
the electrostatics in this periodic space is obtained by summation of
the charge interactions over periodically replicated simulation cells.
This yields the well-known formula of Ewald \cite{Ewald:21} for the
Coulomb energy of charges in a lattice.

Ewald summation is widely used in computer simulations of charged
systems and is generally believed to give the most accurate description
of the electrostatics (with respect to system-size dependence). The
Ewald formula splits the Coulomb energy into two rapidly converging
real- and Fourier-space lattice sums, where their relative
contributions can be controlled by a parameter $\eta$. However, in
numerical implementations one has to apply truncations of the two
(infinite) lattice sums. In this work, we develop a quantitative
description of the errors arising from this truncation. We then discuss
the choice of cutoff distances in real and Fourier space with respect
to numerical accuracy, resulting in restrictions regarding
computational efficiency of Ewald-sum implementations. In particular,
we analyze the connection between the real-space screening parameter
$\eta$ and the number of Fourier-space vectors required for a given
accuracy. We also derive an approximate upper limit for an efficient
choice of the Fourier-space cutoff.

\section{An accuracy measure for truncated Ewald sums}
The Ewald-summation formula for the Coulomb energy $U$ of $m$ charges
$q_\alpha$ (with net charge 0) at positions ${\bf r}_\alpha$ can be
expressed as a sum over pair interactions and self terms,
\begin{eqnarray}
U&=&\sum_{1\leq\alpha<\beta\leq m} q_\alpha q_\beta
\phi({\bf r}_{\alpha\beta})
+\frac{1}{2}\sum_{\alpha=1}^{m}q_\alpha^2 \lim_{{\bf
r}\rightarrow 0}\left[ \phi({\bf r}) - \frac{1}{|{\bf r}|}\right]~,
\nonumber\\
\label{eq:UEw}
\end{eqnarray}
where ${\bf r}_{\alpha\beta}={\bf r}_{\beta}-{\bf r}_{\alpha}+{\bf n}$,
with the lattice vector ${\bf n}$ chosen such that ${\bf
r}_{\alpha\beta}$ is a vector in the unit cell.
This result holds for a background dielectric constant
$\epsilon=\infty$ (or vanishing dipole moment of the cell), as
discussed in ref.~\cite{deLeeuw:80:a}.
The effective pair interaction has the following form,
\begin{eqnarray}
\phi({\bf r})&=&\sum_{\bf n}\frac{{\rm erfc}(\eta|{\bf r}+{\bf n}|)}
{|{\bf r}+{\bf n}|}
+ \sum_{\scriptstyle {\bf k} \atop \scriptstyle k>0}
\frac{4\pi}{V k^2} \exp ( -k^2/4\eta^2 + i{\bf k}\cdot{\bf r})
-\frac{\pi}{V \eta^2}~,
\label{eq:phi}
\end{eqnarray}
where $V$ is the volume of the box, erfc is the complementary error
function, and $k=|{\bf k}|$. The two lattice sums extend over real and
reciprocal (Fourier) space lattice vectors ${\bf n}$ and ${\bf k}$,
respectively. In most practical applications, the screening parameter
$\eta$ is chosen such that only $n=0$ and a few hundred ${\bf
k}$ vectors
need to be considered. The convergence parameter $\eta$ can be chosen
arbitrarily ($0<\eta<\infty$), as
\begin{eqnarray}
\frac{\partial U}{\partial \eta} & \equiv & 0~.
\label{eq:dudeta}
\end{eqnarray}
The choice of $\eta$ gives different weights to the two sums.
However, the requirement of
numerical accuracy imposes some restrictions on the choice of $\eta$.
Any truncation of the two lattice sums in Eq.~(\ref{eq:phi}) results in
deviations from the identity Eq.~(\ref{eq:dudeta}). This provides us
with a measure for the accuracy of a given implementation characterized
by a screening parameter $\eta$ and two cutoff distances $N$ and $K$
for the real- and Fourier-space lattice sums ($N\geq|{\bf n}|$,
$K\geq|{\bf k}|$).

The errors $\Delta U$ of the total energy and $\Delta u_{\alpha}$ of
the single-particle energies are weighted sums of numerical errors
$\Delta\psi$ in the electrostatic interaction $\psi({\bf r})=\phi({\bf
r})-1/r$,
\begin{eqnarray}
\Delta U & = & \left| \frac{1}{2} \sum_{\alpha,\beta=1}^{m}
q_\alpha q_\beta \Delta\psi({\bf r}_{\alpha\beta}) \right| \leq
\frac{1}{2} \max_{{\bf r}\in V}|\Delta\psi({\bf r})|
\left( \sum_{\alpha=1}^{m} |q_\alpha| \right)^2~,
\label{eq:utot-mdep}\\
\Delta u_{\alpha} & = & \left| q_\alpha \sum_{\beta=1}^{m} q_\beta
\Delta\psi({\bf r}_{\alpha\beta}) \right| \leq \left| q_\alpha \right|
\max_{{\bf r}\in V}|\Delta\psi({\bf r})| \sum_{\beta=1}^{m} |q_\beta|~.
\label{eq:using-mdep}
\end{eqnarray}
In most practical cases, we expect a considerably smaller error than
indicated by the upper bounds of Eqs.~(\ref{eq:utot-mdep}) and
(\ref{eq:using-mdep}). A detailed analysis of the errors in disordered
charge configurations has been presented by Kolafa and Perram
\cite{Kolafa:92}. In this work, we focus on a model- and
configuration-independent measure of the numerical accuracy of
truncated Ewald sums. This general error analysis provides insight into
the influence of the Ewald-sum parameters $\eta$, $N$, and $K$ (and,
possibly, a real-space cutoff $R_c$). We do not analyze effects of
partial error cancellation owing to, e.g., charge ordering.

We will study a system of two particles with charges
$\pm 1$, at positions $r=0$ and ${\bf r}$.
We will restrict our analysis to the most widely used cubic cell.
Calculations will be done in reduced coordinates with a
lattice constant of 1, resulting in ${\bf n}$ assuming integer values
and ${\bf k}=2\pi{\bf n}$. We define $\Delta(\eta,N,K)$ as the maximum
deviation from the identity Eq.~(\ref{eq:dudeta}) with respect to ${\bf
r}$ vectors in the cell $V$. $\Delta(\eta,N,K)$
will serve as our measure for the numerical accuracy of a truncated
Ewald sum. From Eq.~(\ref{eq:UEw}) we obtain
\begin{eqnarray}
\Delta(\eta,N,K) = \max_{{\bf r}\in V}\left|
-\frac{\partial U}{\partial\eta}\right|
& = &\max_{{\bf r}\in V}\left|
\sum_{\scriptstyle {\bf n} \atop \scriptstyle n\leq N}
\frac{2}{\pi^{1/2}}\left[
\exp(-\eta^2 n^2)-\exp(-\eta^2|{\bf r}+{\bf n}|^2)
\right]
\right.
\nonumber\\
&&-\left.
\sum_{\scriptstyle {\bf k} \atop \scriptstyle 0<k\leq K}
\frac{2\pi}{\eta^3}\exp(-k^2/4\eta^2)\left[1-
\exp(i{\bf k}\cdot{\bf r})\right]\right|~.
\label{eq:Delta}
\end{eqnarray}
Using the identity Eq.~(\ref{eq:dudeta}) for the full Ewald sum
($N,K\rightarrow\infty$),
we can invert the sign and sum over the complementary ${\bf n}$- and
${\bf k}$-space regions $n>N$ and $k>K$, respectively.
We approximate the $k$-space contributions neglected in
Eq.~(\ref{eq:Delta}) by an integral,
\begin{eqnarray}
\lefteqn{\Delta_k(\eta,K) = \max_{{\bf r}\in V}\left\{
\sum_{\scriptstyle {\bf k} \atop \scriptstyle k>K}
\frac{2\pi}{\eta^3}\exp(-k^2/4\eta^2)\left[1-
\exp(i{\bf k}\cdot{\bf r})\right]\right\}}\nonumber\\
&& \approx \max_{{\bf r}\in V}\left\{
\frac{1}{\pi\eta^3}\int_{K}^{\infty}dk\,k^2\,
\exp(-k^2/4\eta^2)\left[1-\frac{\sin(kr)}{kr}\right]\right\}~,
\label{eq:DeltaInt}
\end{eqnarray}
noting that $\Delta_k\geq 0$.
This results in an approximate expression for the neglected $k$-space
contributions,
\begin{eqnarray}
\Delta_k(\eta,K) & \approx & \sigma(K/\eta)~.
\label{eq:DeltaK}
\end{eqnarray}
We obtain $\sigma(x)$ by integration of Eq.~(\ref{eq:DeltaInt})
replacing $1-\sin(x)/x$ by its maximum value $\gamma$
($\approx 1.22)$,
\begin{eqnarray}
\sigma(x) & = & 2\gamma x\pi^{-1}\exp(-x^2/4) +
2\gamma\pi^{-1/2}{\rm erfc}(x/2)~.
\label{eq:sigma}
\end{eqnarray}
Typically, $\eta$ is large ($\eta>4$) and only one simulation cell is
considered in $r$ space ($N=0$).
Regarding the sign of the real-space contributions for $N=0$,
\begin{eqnarray}
\Delta_r(\eta,N=0,K,{\bf r}) & = &
2\pi^{-1/2} \sum_{\scriptstyle {\bf n} \atop \scriptstyle n>0}
\left[\exp(-\eta^2|{\bf r}+{\bf n}|^2)-\exp(-\eta^2 n^2)\right]~,
\label{eq:Deltar}
\end{eqnarray}
$\Delta_r$ is found to be positive in extensive numerical tests.
Expressed in terms of theta functions using Jacobi's imaginary
transformation \cite{Whittaker:78}, we conjecture
\begin{eqnarray}
0\geq \pi^{3/2}\eta^{-3}\left[
\vartheta_3(\pi x, q) \vartheta_3(\pi y, q) \vartheta_3(\pi z, q)
- \vartheta_3^3(0, q)\right] \geq \exp[-\eta^2 ( x^2 + y^2 + z^2 )]
- 1~,
\label{eq:ineq}
\end{eqnarray}
where $\vartheta_3(u,q)=1+2\sum_{n=1}^{\infty}q^{n^2}\cos(2nu)$,
$q=\exp(-\pi^2/\eta^2)$, and $(x,y,z)\in[-1/2,1/2]^3$. The left part of
the inequality is trivial. However, we could not find a formal proof
for the lower bound. The neglected $r$-space
contributions in $\Delta$ come mainly from the neighboring image and
reach a maximum at the center of the faces of the cube [${\bf
r}=(1/2,0,0)$]. From Eq.~(\ref{eq:Deltar}), the real-space
contributions to $\Delta$ are estimated as
$2\pi^{-1/2}\exp(-\eta^2/4)$, resulting in an approximate expression
for $\Delta$,
\begin{eqnarray}
\Delta(\eta,N=0,K)&\approx& \sigma(K/\eta)
+2\pi^{-1/2}\exp(-\eta^2/4)~.
\label{eq:DeltaApp}
\end{eqnarray}
For the practically less interesting case of $N=0$ and $\eta$
small ($\eta<1$), Eq.~(\ref{eq:ineq}) can be used to derive an
upper bound $\Delta_r\leq 2\pi^{-1/2}[1-\exp(-3\eta^2/4)]$.
Based on an analysis of the dielectric properties of polar fluids in
periodic space, Neumann and Steinhauser \cite{Neumann:83:a} derived a
measure for the effect of a real-space truncation of Ewald sums, which
for $N=0$ gives $Q=[{\rm erf}(\eta/2)]^3$. For large $\eta$, the
deviation from ideality $\delta=1-Q^{1/3}$ scales as $\delta\sim
\exp(-\eta^2/4)/\eta$, similar to what we find based on our analysis of
$\partial U/\partial\eta$.

The dependence of $\Delta$ on the $k$-space cutoff $K$ is depicted in
Fig.~\ref{fig:DeltaMax}. The curves obtained from the approximation
Eq.~(\ref{eq:DeltaApp}) are compared with the maxima calculated from
$10^4$ points ${\bf r}$ randomly chosen in a cubic cell. We observe
excellent agreement of the approximate formula in the range considered
($3\leq\eta\leq 10; 0<K/2\pi\leq 7$). Eq.~(\ref{eq:DeltaApp}) can
therefore be used for assessing the quality of an implementation ($N,
K, \eta$) of the Ewald sum. An interesting observation from
Fig.~\ref{fig:DeltaMax} is that certain $K$ values show a particularly
small Fourier-space errors for all $\eta$ values studied. The $K$ values
are characterized by a gap in the lattice, i.e., there do not exist
${\bf k}$ vectors such that $k^2=K^2+4\pi^2$. The numerical values are
of the form $(K/2\pi)^2=6+8m=6$, 14, 22, 30, 38, and 46
\cite{Mordell:69}; although followed by a gap, $(K/2\pi)^2=27$ does not
give an optimal cutoff.

The inverse of Eq.~(\ref{eq:sigma}) can
be used to obtain the ratio $K/\eta$ given an error in the
$k$-space sum,
\begin{eqnarray}
x(\sigma) & \approx & 2 (-\ln\sigma)^{1/2}
+\ln[4\gamma(-\ln\sigma)^{1/2}\pi^{-1}]\;
(-\ln\sigma)^{-1/2}~,
\label{eq:siginv}
\end{eqnarray}
which is asymptotically correct for small errors $\sigma$, but is
already a good approximation for $\sigma<0.1$.

An important observation is that only ratios $K/\eta$ enter the formula
for the $k$-space error Eq.~(\ref{eq:DeltaK}). Correspondingly, to
achieve the same accuracy with two values $\eta_1$ and $\eta_2$, the
$k$-space cutoff distances have to be chosen proportionally,
$K_1/K_2=\eta_1/\eta_2$. The number of ${\bf k}$ vectors $\nu(K)$ for a
given value of $K$ scales as $K^3$. Thus, the number of ${\bf k}$
vectors required to maintain a given accuracy increases with the third
power of the $\eta$ ratios when increasing $\eta$,
\begin{eqnarray}
\nu(K_1)/\nu(K_2)&=&(\eta_1/\eta_2)^3~.
\end{eqnarray}
 From the analysis of the ${\bf k}$-dependent dielectric constant,
Neumann \cite{Neumann:86:a} proposed a measure $p=3\exp(-K^2/4\eta^2)$
for the Fourier-space error, which also depends only on $K/\eta$ and
agrees closely with the asymptotic behavior of $\Delta_k\sim
\exp(-K^2/4\eta^2)K/\eta$.

We now determine a maximum useful value of $K$, given $N=0$ and $\eta$.
This is obtained from a relation $K(\eta)$, for which the errors of
$k$-space and real-space truncations are equal, such that a further
increase in the number of ${\bf k}$ vectors would not significantly
reduce the overall error. Equating the expressions for the real- and
Fourier-space errors in a cubic lattice, $\sigma(K/\eta) = 2 \pi^{-1/2}
\exp(-\eta^2/4)$, we obtain an approximate expression,
\begin{eqnarray}
K(\eta) & \approx & \eta^2 + \ln ( \eta^2 \gamma^2 / \pi )~,
\label{eq:Keta}
\end{eqnarray}
asymptotically valid for large $\eta$. (The relative errors of
Eq.~(\ref{eq:Keta}) are less than 0.05 and 0.01 for $\eta=3$ and 5,
respectively.)

In many calculations, a spherical real-space cutoff $R_c$
is introduced in Eq.~(\ref{eq:phi}),
i.e., the argument of the ${\bf n}$
sum is multiplied with a unit step function $\Theta(R_c-|{\bf r}+{\bf
n}|)$. To find an approximate expression for
the numerical error of an implementation $(N=0,\eta,K,R_c)$ where
$R_c$ is smaller than half of the box length, we
use Eq.~(\ref{eq:Delta}) modified by a $\Theta$ function.
We approximate the additional real-space contributions to $\Delta$ as
$2\pi^{-1/2}\exp(-R_c^2\eta^2)$, which yields
\begin{eqnarray}
\Delta(\eta,N=0,K,R_c\leq 0.5)
&\approx& \sigma(K/\eta)
+2\pi^{-1/2}\left[ \exp(-\eta^2/4)+
\exp(-R_c^2\eta^2) \right]~,
\label{eq:DeltaAppRc}
\end{eqnarray}
analogous to Eq.~(\ref{eq:DeltaApp}). Typically, $\eta$ is large
and $R_c$ is chosen smaller than 0.5, such that the
$\exp(-R_c^2\eta^2)$ term dominates. We can then invert
Eq.~(\ref{eq:DeltaAppRc}) to find a
generalization of Eq.~(\ref{eq:Keta}). This gives the $k$-space cutoff
$K$ at which real- and Fourier-space errors are approximately equal,
\begin{eqnarray}
K(\eta,R_c) & \approx & 2 R_c \eta^2 + R_c^{-1}
\ln ( 2 \eta \gamma R_c \pi^{-1/2} )~.
\label{eq:KetaRc}
\end{eqnarray}
For $\eta=10$, we find good agreement for $0.2<R_c<0.5$, with the
relative error of Eq.~(\ref{eq:KetaRc}) smaller than 2\%.

\section{Illustrative examples}
We illustrate our error analysis of Ewald sums using a study of Kusalik
\cite{Kusalik:90}, who reports relative errors of dipole-dipole
energies for several configurations of a dipolar soft-sphere fluid
calculated with $N=0$, $4<\eta<7.5$, and $(K/2\pi)^2=22, 30,$ and 42.
Fig.~14 of ref.~\cite{Kusalik:90} shows the relative errors (including
the sign) for the three $K$ values as a function of $\eta$. Given
$K$, the relative errors are minimal for some values of $\eta$.
The optimal combinations of $\eta$ and the $k$-space cutoff
$K$ from Kusalik's calculations are approximately $\eta=5, 5.5$, and 6
for $(K/2\pi)^2=22, 30,$ and 42. These values are in excellent
agreement with those derived from our analysis, with Eq.~(\ref{eq:Keta})
giving $\eta=5.2, 5.6,$ and 6.1 for Kusalik's $K$ values.

In an extension of this study, Kusalik \cite{Kusalik:91:a} reported
electrostatic energy, pressure, and dielectric constant of a dipolar
soft-sphere system for various Ewald-summation parameters ($\eta=10$,
$N=0$, $0.236\leq R_c\leq 0.31$, $46\leq (K/2\pi)^2\leq 82$). However,
the statistical errors--although small--do not allow to establish a
conclusive picture, since all data are approximately within two
estimated standard deviations. Large statistical uncertainties of the
order of 10--20\% also prohibit a detailed examination of the errors of
the dielectric constant of SPC water, as calculated by Belhadj et
al. \cite{Belhadj:91}.

To further demonstrate the quantitative power of the proposed error
analysis, we have studied the energies of random configurations of $m$
charges $\pm 1$ in a cubic box (with net charge 0). Energies have been
calculated for 10 configurations with $m=8$ and $m=32$ point charges
using Ewald summation ($N=0,K,\eta)$ and by explicit lattice sums using
lattice vectors ${\bf n}$ with $|{\bf n}|\leq 50$ (with the correction
for a net dipole moment of the box considered \cite{deLeeuw:80:a}). The
relative errors $\rho$ in the energy with respect to the lattice sums
have been determined for $(K/2\pi)^2=10,20,\ldots,90$ and
$3\leq\eta\leq 10$. For given values of $K$, we determine the screening
parameter $\eta$ such that the relative errors $\rho(K,\eta)$ assume a
minimum.

Fig.~\ref{fig:Keta} shows the relation between optimal $K$ and $\eta$
values together with the derived curve $K(\eta)$ from
Eq.~(\ref{eq:Keta}). We observe excellent qualitative agreement between
the derived relation $K(\eta)$ and the observed minima.
Quantitatively, the results for the random configurations suggest
somewhat larger $k$-space cutoff distances $K$ for given $\eta$.
However, in most practical applications the charges are more
effectively screened by neighboring charges than in random
configurations, such that the $k$-space contributions to the energy
tend to be smaller, justifying somewhat smaller $K$ values.

Another important point is the relation between $\Delta$ and the
relative errors $\rho(K,\eta)$ in the energy. For $\rho(K,\eta)$, the
results for the random configurations have been used. $\Delta$ has been
calculated from Eqs.~(\ref{eq:DeltaApp}) and (\ref{eq:Keta}).
Fig.~\ref{fig:rhok} shows minimum values (for given $K$) of
$\rho(K,\eta)$ and $\Delta$ as a function of $K$. $\rho$ and $\Delta$
closely follow each other, supporting the present error analysis. For
the random configurations, they are proportionally related with a
factor of about 100.

\section{Discussion}
The present error analysis has important implications on the choice of
the Ewald-sum parameters $\eta$, $K$, and $R_c$ in computer simulations
of condensed-matter systems, helping to avoid unnecessary computational
effort and minimize the numerical error. The analysis of truncation
errors allows to choose $\eta$, $K$, and $R_c$ on a rational basis.
Using the accuracy measure $\Delta$, it becomes possible (i) to
assess the numerical quality of an Ewald-sum implementation and (ii)
to compare different implementations using different parameters.

An important application of the Ewald-summation error analysis in
computer-simulation studies is to optimize the choice of the screening
parameter $\eta$, the real-space cutoff $R_c$, and the Fourier-space
cutoff $K$ regarding computational speed
\cite{Kolafa:92,Perram:88,Fincham:94,Rycerz:92:b}. Using a few typical
configurations of the system, one can minimize the computer time for
the energy (or force) calculation using combinations of $\eta$ and $K$
that give the same error $\Delta$. The inversion of
Eqs.~(\ref{eq:DeltaApp}) and (\ref{eq:DeltaAppRc}) yields the
appropriate expressions for $K$,
\begin{eqnarray}
K(\eta,\Delta)&=&\eta\;\sigma^{-1}\!\left[
\Delta - 2\pi^{-1/2}\exp(-\eta^2/4)\right]~,\\
K(\eta,\Delta,R_c)&=&\eta\;\sigma^{-1}\!\left\{
\Delta - 2\pi^{-1/2}\left[\exp(-\eta^2/4)
+\exp(-\eta^2 R_c^2)\right]\right\}~,
\end{eqnarray}
where $\sigma^{-1}$ is the inverse of $\sigma$, as defined in
Eq.~(\ref{eq:sigma}). An approximate analytical expression for
$\sigma^{-1}$ is given in Eq.~(\ref{eq:siginv}). This strategy of
optimizing $(\eta,K)$ pairs along curves of constant error is
particularly important in computational studies of large Coulombic
systems (e.g., biomolecules in solution), where an overwhelming amount
of computer time is spent on the calculation of long-range charge
interactions.

\section*{Acknowledgment}
The author wants to thank A. E. Garc\'{\i}a and M. Neumann for
many helpful discussions. This work has been funded by the
Department of Energy (U.S.).

\begin{figure}
\caption{Error $\Delta$ of an Ewald-sum implementation ($N=0,\eta,K$)
on a logarithmic scale
as a function of the $k$-space cutoff $K/2\pi$. The symbols show
the maximum value of
$\Delta$ [Eq.~(\protect\ref{eq:Delta})]
in the cell
estimated from $10^4$ random points. The lines are calculated
from the approximate expression Eq.~(\protect\ref{eq:DeltaApp}).
Results for $3\leq\eta\leq 10$ are shown.}
\label{fig:DeltaMax}
\end{figure}
\begin{figure}
\caption{Optimal combination of screening parameter $\eta$ and
$k$-space cutoff $K$. The symbols indicate average values of $\eta$
that minimize the relative error $\rho(K,\eta)$ in the energy.
$\rho(K,\eta)$ has been calculated for 10 random configurations of
$m=8$ ($\diamond$) and $m=32$ ($+$) charges $\pm 1$ in
periodically-replicated cubic boxes. The solid line is the predicted
result from Eq.~(\protect\ref{eq:Keta}).}
\label{fig:Keta}
\end{figure}
\begin{figure}
\caption{Relative errors $\rho(K,\eta)$ of the energy and Ewald-sum
accuracy measure $\Delta$ for optimal combinations of screening
parameter $\eta$ and $k$-space cutoff $K$. Relative errors
$\rho(K,\eta)$ have been calculated for random configurations of
$m=8$ ($\diamond$) and $m=32$ ($+$) charges $\pm 1$;
they are compared with $\Delta$ (-----)
calculated from Eqs.~(\protect\ref{eq:DeltaApp}) and
(\protect\ref{eq:Keta}).}
\label{fig:rhok}
\end{figure}
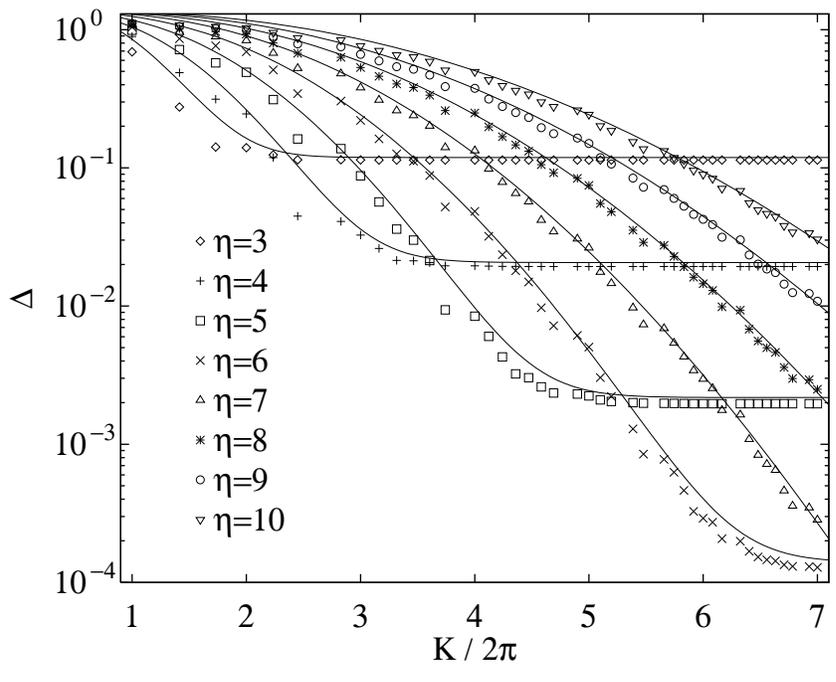
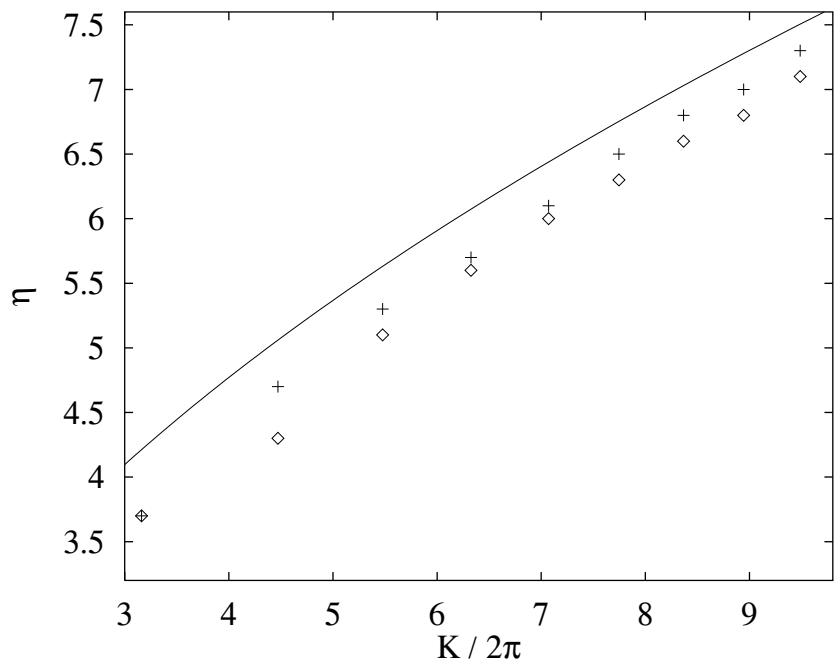
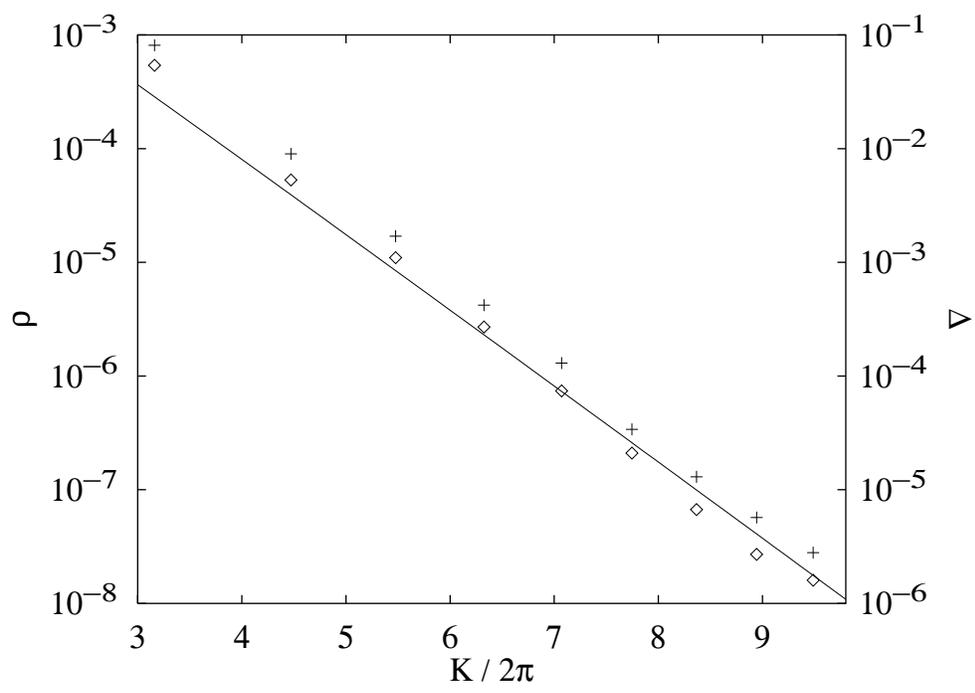
\end{document}